# A possible bright ultraviolet flash from a galaxy at redshift *z* ≈ 11

Linhua Jiang[1,2], Shu Wang[1,2], Bing Zhang[3], Nobunari Kashikawa[4,5], Luis C. Ho[1,2], Zheng Cai[6], Eiichi Egami[7], Gregory Walth[8], Yi-Si Yang[9,10], Bin-Bin Zhang[9,10], Hai-Bin Zhao[11]

*[1]Kavli Institute for Astronomy and Astrophysics, Peking University, Beijing, China*

*[2]Department of Astronomy, School of Physics, Peking University, Beijing, China*

*[3]Department of Physics and Astronomy, University of Nevada, Las Vegas, NV, USA*

*[4]Department of Astronomy, Graduate School of Science, The University of Tokyo, Tokyo, Japan*

*[5]Optical and Infrared Astronomy Division, National Astronomical Observatory, Tokyo, Japan*

*[6]Department of Astronomy, Tsinghua University, Beijing, China*

*[7]Steward Observatory, University of Arizona, Tucson, AZ, USA*

*[8]Observatories of the Carnegie Institution for Science, Pasadena, CA, USA*

*[9]School of Astronomy and Space Science, Nanjing University, Nanjing, China*

*[10]Key Laboratory of Modern Astronomy and Astrophysics (Nanjing University), Ministry of Education, China*

*[11]CAS Key Laboratory of Planetary Sciences, Purple Mountain Observatory, Chinese Academy of Sciences, Nanjing, China*

**In the optical sky, minutes-duration transients from cosmological distances are rare. Known objects that give rise to such transients include gamma-ray (γ-ray) bursts (GRBs), the most luminous explosions in the universe[1] that have been detected at redshift as high as $z \sim 9.4$ (refs. 2-4). These high-redshift GRBs and their associated emission can be used to probe the star formation and reionization history in the era of cosmic dawn. Here we report a near-infrared (near-IR) transient with an observed duration shorter than 245 s coincident with the luminous star-forming galaxy GN-z11 at $z \approx 11$. The telluric absorption shown in the near-IR spectrum indicates its origin from above the atmosphere. We can rule out the possibility of known man-made objects or moving objects in the Solar system based on the observational information and our current understanding of the properties of these objects. Since some long-duration GRBs are associated with a bright ultraviolet (UV) or optical flash[5-14], we investigate the possibility that the detected signal arose from a rest-frame UV flash associated with a long GRB from GN-z11. Despite the very low probability of being a GRB, we find that the spectrum, brightness, and duration of the transient are consistent with such an interpretation. Our result may suggest that long GRBs can be produced as early as 420 million years after the Big Bang.**

GN-z11 is a luminous star-forming galaxy at redshift $z \approx 11$. It was photometrically selected as a candidate galaxy at $z > 10$ in the Hubble Space Telescope (HST) GOODS-North field. GN-z11 has a magnitude of 26.0 in the HST Wide Field Camera 3 (WFC3) F160W band (magnitudes are on the AB system). Follow-up WFC3 near-IR grism observations suggested a redshift of $z = 11.1 \pm 0.1$ (ref. 15). On 2017 April 7 (universal time), we carried out $K$-band spectroscopic observations of GN-z11 using the near-IR, multi-object spectrograph MOSFIRE[16] on the Keck I telescope. Our slit mask included GN-z11 in the middle, along with two bright reference stars and 18 low-redshift galaxies. The slit width was 0.9 arcseconds ("). The spectroscopy was made under clear observing conditions with ~0.6 - 0.7" seeing. We obtained a total of 5.3 h of on-source integration, consisting of 106 individual 179 s exposures. The gap between two consecutive exposures was 33 s. From the combined $K$-band spectrum, we plausibly detected three rest-frame UV emission lines and determined a spectroscopic redshift $z = 10.957 \pm 0.001$ for GN-z11. See Methods and our companion paper[17] for details.

In one of the 106 individual $K$-band images, we detected a bright burst: a compact continuum emission with a spatial position coincident ($\pm 1$ pixel) with the position of the GN-z11 UV emission lines. This burst is hereafter referred to as GN-z11-flash. Fig. 1 shows part of the two-dimensional (2D) image (ID209) that contains GN-z11-flash, as well as four images taken immediately before and after it. The full 2D spectrum in the $K$ band is shown in Fig. 2a. A direct extraction of GN-z11-flash is displayed as the one-dimensional (1D) spectrum in Fig. 2b. It exhibits prominent telluric absorption features, indicating that GN-z11-flash arose from above the atmosphere. After correcting for telluric absorption and applying flux calibration, we obtained the calibrated spectrum shown in Fig. 2c. The spectrum is fit using a power law with a slope $\beta_\lambda = -3.2 \pm 0.4$ ($f_\lambda \propto \lambda^\beta$) or $\beta_\nu = 1.2 \pm 0.4$ ($f_\nu \propto \nu^\beta$), where $f$ is flux density, $\lambda$ is wavelength, and $\nu$ is frequency. The spectrum does not show noticeable emission or absorption features. The flux density in the $K$ band (effective wavelength ~ 2.15 μm) calculated from this observed spectrum is 0.057 mJy, corresponding to 19.5 mag. The flux densities and luminosities herein are time-averaged values over the 179 s exposure.

We have performed a comprehensive analysis of the origin of GN-z11-flash based on the observational information (date, time, site), target information (R.A. and Dec.), slit mask design (slit position angle, target position), and the properties (orbit, brightness, and angular velocity) of man-made satellites and moving objects in the Solar system (Methods). First, GN-z11-flash was very unlikely to be caused by a ghost image or a diffraction spike of a bright star, because there are no bright stars in the field or in its ambient region. Second, the $K$-band spectrum with a blue continuum strongly constrains the source properties and rules out monochromatic light such as a laser or airplane strobe light that has strong emission lines. Third, we will argue that GN-z11-flash was highly unlikely from a moving object. A moving object would have produced an extended spectral shape and likely similar spectra in more than one slit in different images, unless its trajectory was nearly perpendicular to the GN-z11 slit. In addition, a moving object must be bright enough (e.g., a naked-eye low-Earth orbit satellite) to produce the observed spectrum, because the time duration across the 0.9"-wide slit was short.

We rule out the possibility that GN-z11-flash was from a man-made satellite. There are three classes of satellite orbits: low-Earth, medium-Earth, and high-Earth. As satellites reflect sunlight, low-Earth orbit satellites are nearly invisible during the night. GN-z11-flash was detected 3.5 hours after sunset, and GN-z11 was rising in the eastern sky. At that moment for an observer at the observing site (Mauna Kea at latitude ~20º north) toward the direction of GN-z11, a low-Earth or medium-Earth orbit satellite with height ≲4000 km would not have been able to reflect the light of the Sun. However, a satellite above ~4000 km would have an inclination angle ≳38˚ (the angle between the satellite orbit and the Earth's equator), which would produce an extended spectrum that is inconsistent with the compact spectrum shown in Fig. 1 (see also Extended Data Fig. 1). For high-Earth orbit (≳36,000 km) satellites over the Earth's equator, Mauna Kea observers should not see them toward the direction of GN-z11. We have also checked public satellite databases and found no catalogued satellites in tandem.

We checked the IAU Minor Planet Center database and rule out the possibility that GN-z11-flash was from a known moving object in the Solar system. The majority of the known asteroids orbit in or near the plane of the Ecliptic, but GN-z11 is far from this plane. The chance probability that we caught a bright, unknown asteroid (or minor planet) is of order of $10^{-13}$ during our observations. We also rule out the possibility of a foreground brown dwarf flare aligned with GN-z11, because it is not consistent with the deep HST and Spitzer photometry[15]. Finally, the short duration of the $K$-band flash is hard to explain within the framework of transients other than GRBs, such as supernovae (SNe) or tidal disruption events (TDEs). The chance probability of having a foreground GRB flash coincident with GN-z11-flash is extremely low. We estimate this probability using a solid angle of 0.9" × 0.7", where 0.9" is the slit width and 0.7" is our tolerance of the position uncertainty on the slit (roughly four pixels). This solid angle extends a fraction of ~$1.2 \times 10^{-12}$ of the whole sky. Considering an all-sky GRB event rate of ~ 3 per day, each having a $K$-band flash lasting for ~300 s, the probability of having a foreground GRB flash in the solid angle is only ~ $1.3 \times 10^{-14}$. Therefore, we conclude that GN-z11-flash likely originated from the GN-z11 host galaxy.

Adopting a redshift of $z$ = 10.957 for GN-z11-flash and assuming $H_0 = 68$ km s$^{-1}$ Mpc$^{-1}$, $\Omega_m = 0.3$, and $\Omega_\Lambda = 0.7$ leads to an absolute magnitude of −28.2 at rest-frame 1800 Å (1.67 × $10^{15}$ Hz), corresponding to a luminosity of $\nu L_\nu = \lambda L_\lambda = 1.34 \times 10^{47}$ erg s$^{-1}$. This is a lower limit. Our best estimate of the flux density, corrected for light loss, is ~0.1-0.2 mJy (Methods), and thus $\nu L_\nu = (2.35−4.70) \times 10^{47}$ erg s$^{-1}$. Such a high luminosity is consistent with being associated with a long-duration GRB, but outshines other known UV, optical, or infrared transients (e.g., SNe, TDEs, and active galactic nucleus flares). Observations have shown that some long GRBs are associated with a bright UV/optical flash[5-9], which may originate from the reverse shock when the relativistic GRB outflow is decelerated by an ambient medium or in an internal $\gamma$-ray emission site[10-14]. We search for a possible GRB temporally and spatially coincident with GN-z11-flash using archival Fermi Gamma-ray Burst Monitor (GBM) data. The non-detection sets up an upper limit of the $\gamma$-ray flux density $\sim 5.9 \times 10^{-7}$ erg s$^{-1}$ cm$^{-2}$ in the GBM band (Methods), which can be translated to an upper limit on the isotropic luminosity of $\sim 9.8 \times 10^{53}$ erg s$^{-1}$ for the GRB. Most GRBs have an isotropic luminosity below this value, so the non-detection is reasonable.

We explain GN-z11-flash as a rest-frame UV flash associated with a long GRB at GN-z11 (Methods). The hard continuum spectrum with $\beta_\nu \approx 1.2$ suggests that the frequency $\nu$ of the observed emission is slightly lower than the synchrotron self-absorption (SSA) frequency $\nu_a$ (i.e. $\nu \leq \nu_a$). We find that GN-z11-flash was not from an external reverse shock. The most likely solution is that this flash was produced together with the prompt $\gamma$-ray emission from an internal site, plausibly an internal magnetic dissipation site. Based on the moderately fast cooling synchrotron radiation model[18] and the standard method[14] to calculate $\nu_a$, we reach a self-consistent solution of $\nu_a$ with the following parameters for a typical bright long GRB: an isotropic peak luminosity $10^{53}$ erg s$^{-1}$, an observed peak energy 100 keV, a Lorentz factor 300, a co-moving magnetic field strength $10^5$ G, and the distance of an emission region from the central engine $5 \times 10^{14}$ cm. Within this scenario, the duration of the flash is defined by the duration of the burst, which is shorter than $245/(1+z) \approx 20.5$ s in the rest frame. This is consistent with the typical duration $\sim 10$ s of long GRBs[19,20]. Our simulation also shows that $\sim 51\%$ of prompt emission events from long GRBs at $z \sim 11$ would show up in only one image during our observations. Since no obvious continuum radiation was detected after the detection of GN-z11-flash in our observations, we rule out a bright reverse shock component. This is consistent with the GRB models in a wide range of parameter space, including a Poynting-dominated flux or a normal fireball without substantial magnetization in the reverse shock region[21]. For typical GRB parameters, the predicted flux density of the external forward shock emission is below the flux density upper limit set by the non-detections in the observations.

Long GRBs reside in active star-forming galaxies. GN-z11 is a luminous star-forming galaxy with a UV star formation rate of $\sim$26 solar masses per year[17]. During the observations of GN-z11, the chance probability of detecting one GRB as bright as GN-z11 in the UV/optical is estimated to be $(0.3 \sim 60) \times 10^{-10}$ (Methods). This probability is low, but is roughly $10^3$-$10^5$ times higher than the chance probability of detecting a random GRB, and is at least 2 orders of magnitude higher than the probabilities from other sources considered earlier. Our result may also hint at a much higher GRB rate at very high redshifts. Observationally, long GRBs are associated with Type Ibc supernovae[22-26], implying that GRB progenitors have lost both hydrogen and helium envelopes. This is consistent with Wolf-Rayet (WR) progenitor stars. Theoretical modelling[27] has shown that these stars can give birth to long GRBs. We have detected strong [C III] λ1907 and C III] λ1909 emission from GN-z11 that usually requires the presence of a very young stellar population[17]. Such a young population may suggest a large number of WR stars, which further increases the GRB rate. This strengthens the case in support of our hypothesis of a GRB origin for the GN-z11-flash.

The previous redshift record for GRBs was 9.4 (ref. 4). The discovery of GN-z11-flash explained as a GRB-associated UV flash has possibly pushed this record to $z = 10.957$. Since a WR progenitor is preferred for this putative long GRB event, our results suggest that Population II stars have already formed at such a high redshift. The first-generation stars (Population III stars) must have formed at even higher redshifts, consistent with theoretical predictions[28,29]. Our results also suggest that the GRB event rate could be very high in the early universe, implying rapid galaxy formation. More sensitive GRB detectors will be able to directly observe these GRBs in the future[30], and probe the early epoch of cosmic reionization.

Correspondence and requests for materials should be addressed to L.J. (jiangKIAA@pku.edu.cn) or B.Z. (zhang@physics.unlv.edu).

**Acknowledgements** We acknowledge support from the National Science Foundation of China (11721303, 11890693, 11991052), the National Key R&D Program of China (2016YFA0400702, 2016YFA0400703), and the Chinese Academy of Sciences (CAS) through a China-Chile Joint Research Fund (1503) administered by the CAS South America Center for Astronomy. N.K. acknowledges support from the JSPS grant 15H03645. The data presented herein were obtained at the W. M. Keck Observatory, which is operated as a scientific partnership among the California Institute of Technology, the University of California, and the National Aeronautics and Space Administration. The Observatory was made possible by the generous financial support of the W. M. Keck Foundation. We wish to recognize and acknowledge the very significant cultural role and reverence that the summit of Mauna Kea has always had within the indigenous Hawaiian community. We are most fortunate to have the opportunity to conduct observations from this mountain. This research has made use of data provided by CalSky.com and by the International Astronomical Union's Minor Planet Center.

**Author contributions** L.J. designed the program, carried out the Keck observations, analysed the data, and prepared the manuscript. S.W. and G.W. reduced the Keck images. B.Z. proposed the theoretical interpretation and helped to prepare the manuscript. N.K. assisted in the design of the program and carrying out the observations. L.H. helped to prepare the manuscript. Y.S.Y. and B.B.Z. searched the archival Fermi GBM data. H.B.Z. searched the IAU Minor Planet Center database. All authors helped with the scientific interpretations and commented on the manuscript.

**Competing interests** The authors declare no competing interests.

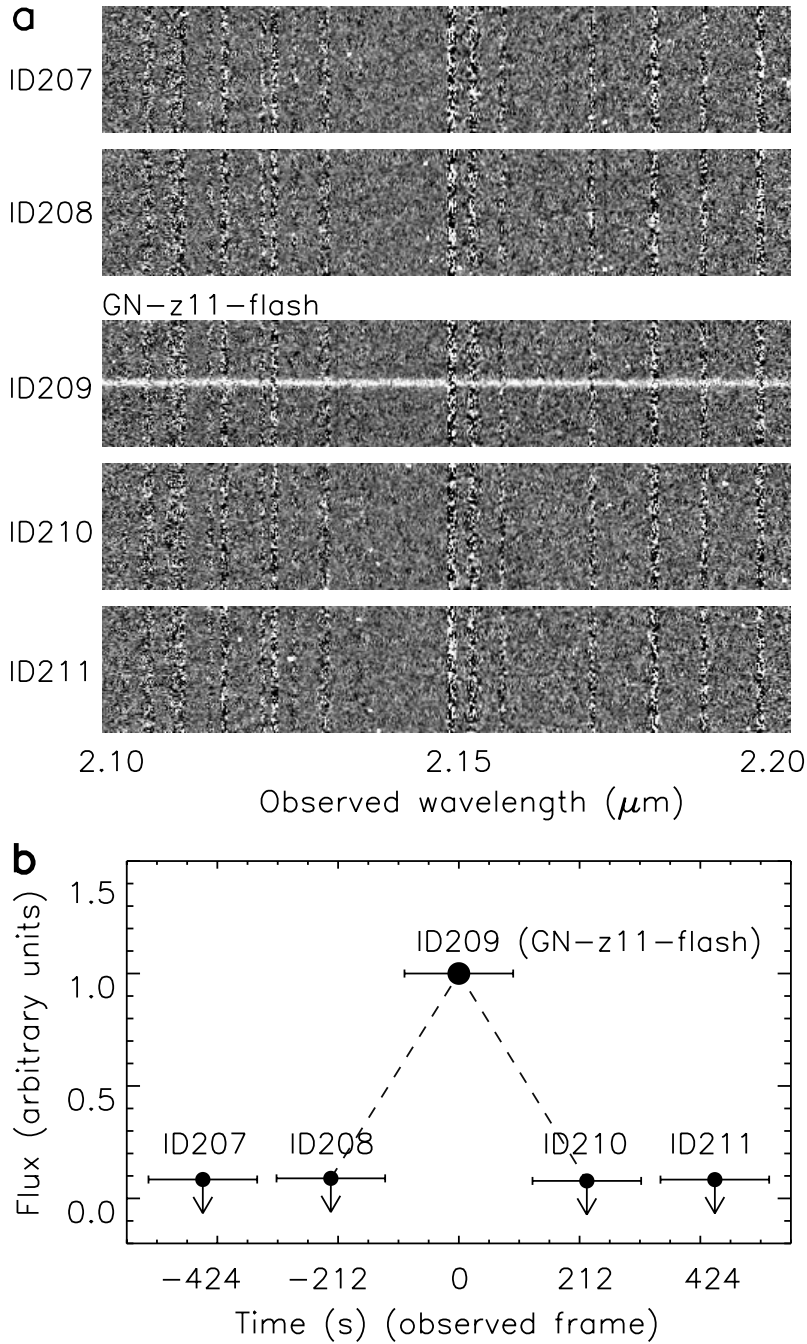

**Fig. 1. Detection of GN-z11-flash.** (a) 2D images that cover a wavelength range of ∼ 2.10 − 2.20 μm. Image ID209 clearly shows a continuum signal that is referred to as GN-z11-flash. The other four images (ID207, ID208, ID210, and ID211) were observed immediately before and after ID209. They do not have detectable signal. (b) Exposure time series. The horizontal bars indicate the individual exposure time of 179 s. The flux of GN-z11-flash in ID209 is normalized to 1 and its error is negligible (the error bar is within the fill circle). The flux measurements in the other four images are shown as 2σ upper limits.

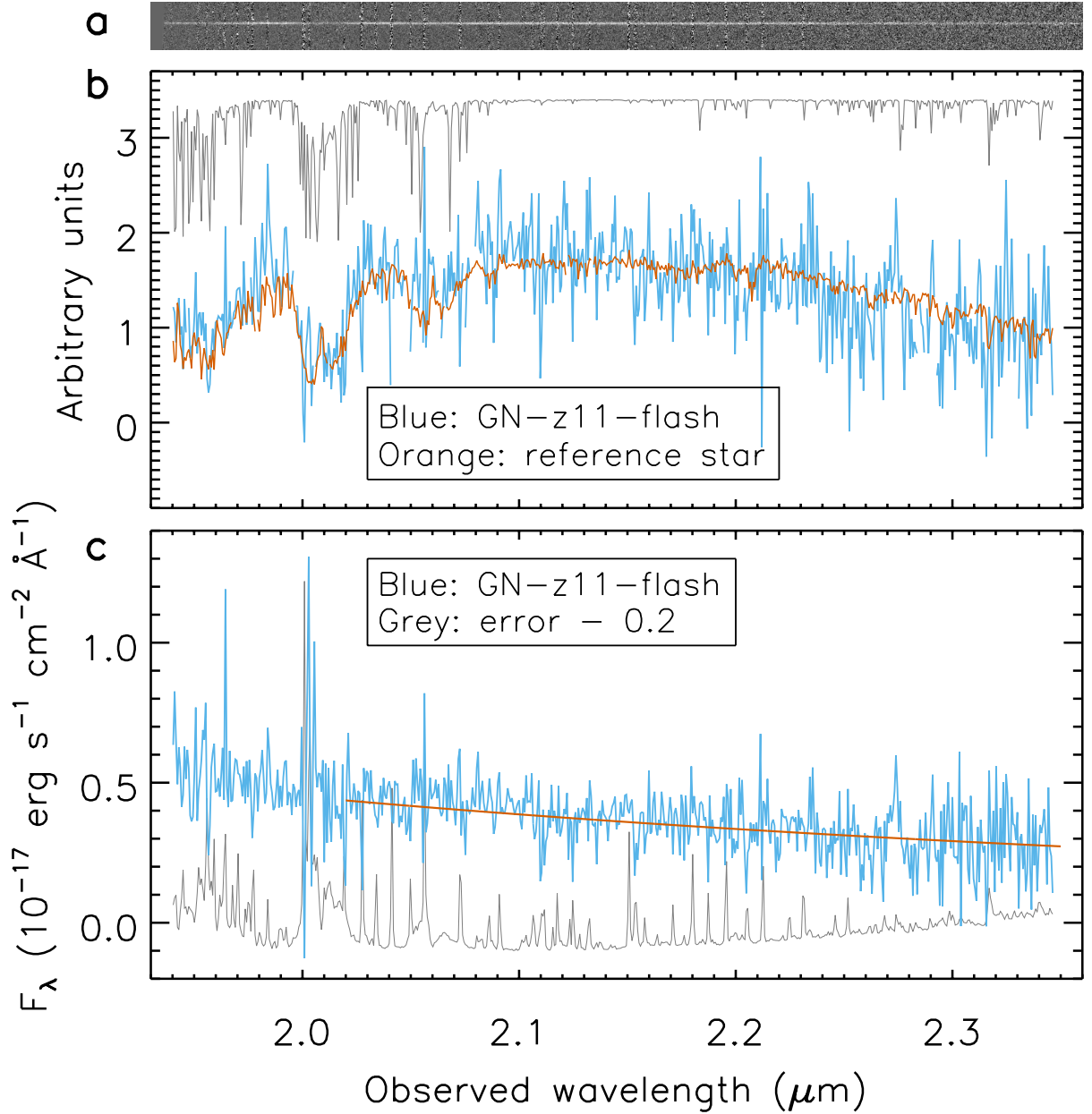

**Fig. 2. Spectra of GN-z11-flash.** (a) The full 2D image of GN-z11-flash in the *K* band. (b) 1D spectrum without calibration. The blue spectrum is a direct extraction of the 2D image in (a). The spectrum of a reference star is scaled and overplotted in orange. A typical sky transparency curve is shown in grey. The spectrum of GN-z11-flash displays strong telluric absorption features, indicating that it arose from above the atmosphere. (c) 1D spectrum after correction for telluric absorption and flux calibration (blue). The orange profile is the best fit of a power law with a slope $\beta_\lambda = -3.2 \pm 0.4$. The 1σ error spectrum in grey is shifted down by 0.2 for clarity.

## Methods

**Keck observations and data reduction.** We carried out $K$-band spectroscopic observations of GN-z11 using Keck MOSFIRE on 2017 April 7 (universal time). MOSFIRE has an effective field-of-view (FoV) of 3×6 arcminutes (′). Our slit mask was designed to place GN-z11 near the center of the field, with a slit position angle (345.5˚) that minimizes light pollution from nearby objects. One bright reference star was chosen for a slit next to that of GN-z11, and it was used for flux calibration. The slit mask also includes 18 targets at lower redshifts. We used the classic ABBA observing mode. The slit width was 0.9", which delivers a resolving power of ~2800. A standard star HIP 68767 (spectral type A0) was observed prior to or after the science observations to correct for telluric absorption. See our companion paper[17] for more information on the Keck observations and data reduction.

**Detection of GN-z11-flash.** We detected a bright burst in one of the individual $K$-band images on 2017 April 7 (UTC 08:07:19.86). It appears as strong continuum emission with a spatial position coincident (±1 pixel) with the position of the GN-z11 UV emission lines. The airmass was 1.45. Figures 1 and 2 illustrate the detection of this event. Fig. 1a also shows the 2D images observed immediately before and after the flash event. The 2σ upper limit of the flux detection in these images is roughly 10% of the flux of GN-z11-flash. We searched the IAU Minor Planet Center database (https://minorplanetcenter.net/) and checked all the raw pointing files of 77 observatories/surveys in the database. We found that the GN-z11 area was not observed on the day of the flash event. In the following paragraphs, we will argue that GN-z11-flash was not from a man-made satellite or a moving object such as an asteroid in the Solar system. Instead, it likely originated from the GN-z11 host galaxy.

First, GN-z11-flash was not an artifact or instrumental effect, or any man-made signal on the Earth's surface. Its spectrum in Fig. 2b unambiguously shows the telluric absorption features, so it must arise from above the atmosphere. In addition, GN-z11-flash was not from a ghost image or a diffraction spike of a bright star, because it is located in the Hubble Deep Field (HDF), which was designed to have no bright stars. In the 10′×10′ region centred on GN-z11, there is only one object slightly brighter than 14 mag in the $r$ band ($K \sim 12$ Vega mag). In the central 15′×15′ region, there are no objects brighter than $r = 12$ mag ($K \sim 10$ Vega mag). Note that our FoV is only 3′×6′. Based on our images with 3-min exposures, an object with $K \sim 10$ Vega mag would nearly saturate some pixels. Also, we did not detect a signal like GN-z11-flash in other slits of the same image or in other images. Therefore, to the best of our knowledge, the probability of GN-z11-flash being a man-made signal on the Earth's surface or a ghost image of a bright star is zero.

Second, the blue continuum emission of GN-z11-flash puts strong constraints on the properties of the source. It can rule out possible man-made monochromatic light from above the atmosphere such as that from a laser and airplane strobe light that has strong emission lines. For example, airplane strobe light is traditionally produced by rare gases and is composed of strong emission lines. In addition, the compact size of the $K$-band spectrum can rule out the possibility of airplane strobe light, because the light needs to be a point source, corresponding to a size of < 4 cm at 10,000 m (or < 2 cm at 5000 m). The size of a light bulb and its associated reflector on airplanes is much larger than this. Therefore, airplane strobe light would have produced an extended spectrum. If GN-z11-flash was from the thermal emission of a man-made object, the temperature of this object should be ≳3000 K in order to produce such a blue continuum slope. To the best of our knowledge, there are no man-made objects with a surface temperature above 3000 K orbiting the Earth. Therefore, the probability from a laser, airplane strobe light, or the thermal emission of a man-made object is zero.

Third, if GN-z11-flash was from a moving object, this object must be bright enough to produce the observed spectrum, because the time duration across the slit was short. If the moving object reflects the light of the Sun, we estimate from its $K$-band magnitude that the $V$-band magnitude of the flash would be roughly 19.2 mag. We will use this magnitude in the $V$ band for the following calculations. The object itself, depending on its angular velocity, could be much brighter than 19.2 mag in order to produce a 19.2 mag signal when it passed across the slit. For example, for a typical low-Earth orbit (~1000 km) satellite with a period of 90 min per orbit, the angular velocity with respect to the centre of the Earth is 4′ per second, or roughly 20′-30′ per second with respect to an observer, depending on the relative positions of the satellite and observer. This requires a $V \sim 5.5$ mag brightness (visible to the naked eye) to produce the observed spectrum when it passes across the 0.9"-wide slit. In addition, if GN-z11-flash was from a moving object, this object would have produced one or more similar spectra in one or more other slits in different images, unless its trajectory was nearly perpendicular to the slit (~ 90˚±20˚). This is because our slit mask contained 21 targets, including GN-z11, two bright reference stars, and 18 low-redshift galaxies. There were no gaps between the slits in the spatial direction, and the GN-z11 slit was placed in the middle of the mask. In fact, the compact spectral shape (Extended Data Fig. 1) already suggests that the trajectory of any hypothetical moving object must be nearly perpendicular to the slit, or else the spectrum would be broader.

Fourth, we ruled out the possibility that GN-z11-flash was from a man-made satellite. If GN-z11-flash was from a moving object, this object was mostly likely a satellite, because we frequently see bright satellite trails in ground-based imaging data. There are three classes of satellites (or orbits): low-Earth orbit, medium-Earth orbit, and high-Earth orbit. The majority of the satellites belong to low-Earth and high-Earth orbits. Satellites reflect the light of the Sun, so low-Earth orbit satellites are invisible or very faint during the night. They are bright only in the dawn or dusk hours. Nearly all high-Earth orbit satellites (≥36,000 km) pass over the Earth's equator. The Mauna Kea site is at latitude ~20º north, and GN-z11 is at Dec. = 62º. We should not see a high-Earth orbit satellite toward GN-z11 at the time of GN-z11-flash. Medium-Earth orbit satellites are rarer compared to the other two classes of satellites. Most of them are navigation satellites and are thus well catalogued.

GN-z11-flash was detected at 10:07 pm local time, roughly 3.5 hours after sunset. GN-z11 was rising in the eastern sky at a low hour angle of about –3 hours. At that moment for an observer at Mauna Kea, a low-Earth or medium-Earth orbit satellite with height ≲4000 km in the direction of GN-z11 would not have been able to directly reflect the light of the Sun. Given the brightness of GN-z11-flash, it should not be from a satellite with height ≲4000 km. On the other hand, a satellite above ~4000 km would have an inclination angle ≳38˚ (the angle of the orbit in relation to the Earth's equator), which would produce an extended spectrum that is inconsistent with the compact spectrum shown in Fig. 1 and Extended Data Fig. 1. A quick calculation is as follows. As shown above, the inclination angle ~38˚ is the minimum angle that we can see a satellite at the moment of GN-z11-flash. The slit position angle was 345.5˚, so the smallest angle between the satellite orbit and the perpendicular direction of the slit is ~23˚. This means that the satellite would move tan (23˚) × 0.9" ~ 0.4" along the slit when it passes across the 0.9"-wide slit. Since the PSF was only ~0.6", the spectrum should be much broader than that obtained. In summary, a satellite below height ~4000 km would not been seen by us, and a satellite above ~4000 km would have produced a wider spectrum. Therefore, to the best of our knowledge, the probability that GN-z11-flash was from a man-made satellite is zero.

We also checked satellite databases and rule out the possibility that GN-z11-flash was from a known satellite. We searched a public satellite database (https://www.calsky.com/) and found one satellite within a distance of 30′ and a duration of 10 minutes at the time of the flash. It was ~ 440 times fainter than required when its angular velocity is considered. In addition, the angle between our slit and the satellite orbit was only 25˚. It would produce a very extended emission, contrary to the narrow shape of GN-z11-flash. There were two satellites within a distance of 3 degrees (greater than 30′), but the angles between their orbits and the slit were only 21˚ and 12˚, respectively. The search result is consistent with the above calculation that bright satellites at the moment of GN-z11-flash should have high inclination angles (or small angles between their orbits and the slit).

Fifth, we checked the IAU Minor Planet Center database and rule out the possibility that GN-z11-flash was from a known moving object in our Solar system. Within a distance of 30′ at the time of the flash, we did not find any known, moving objects, including near-Earth objects, comets, asteroids, and other minor planets. The majority of the known asteroid orbits are in or near the plane of the Ecliptic, but GN-z11 (or HDF) is far from this plane, rendering the surface density of asteroids extremely low. It is highly unlikely that we have caught an unknown, bright asteroid (or minor planet) during our observations. We estimate the chance probability as follows. We assume an angular velocity of ~1" per 3 min (i.e., this object just passed across the slit within one exposure). This is a typical velocity for known asteroids. In the total of 5.3 h, it will cover a solid angle of 106" × 0.36", where 0.36" is the size of two pixels. This solid angle extends a fraction of ~7.2 × 10^{-11} of the whole sky. Based on the above IAU database, about 7400 new moving objects were discovered in 27.8 million images by a few tens of surveys/observatories in 2019, covering about 3/4 of the whole sky. Each piece of the covered sky was observed more than 100 times on average. This means ~20 new objects per day, or 4.5 objects every 5.3 h. If these objects randomly show up in the sky, the chance probability for us to catch a new object within 5.3 h is $7.2 \times 10^{-11} \times 4.5 \times 4/3 \approx 4.3 \times 10^{-10}$. Next, we put three conservative constraints. The first one is that the object trajectory should be nearly perpendicular to the slit. We assume that 10% of the objects satisfy this requirement. The second one is that the majority of asteroids are in or near the plane of the Ecliptic, so the probability that an object shows up in the direction of GN-z11 is much lower. We assume that this probability is 10% of the probability in a random case (the real value could be orders of magnitude lower). The last constraint is that the majority of new/unknown objects are fainter than GN-z11-flash. We assume that 10% of the unknown objects are brighter than GN-z11-flash. After we consider these constraints, the final chance probability is $\sim 4.3 \times 10^{-13}$, which is negligibly small. Note that the assumption of the angular velocity is not important. If the angular velocity is higher, the object would cover a larger solid angle. Meanwhile, this object must be brighter and thus rarer. The two factors roughly cancel out. In principle, one should use the number of all unknown objects (which is clearly unknown) instead of the number of unknown/new objects discovered in one year. But the majority of the newly discovered objects are fainter than GN-z11-flash, and the number of bright objects being discovered is getting smaller. If we increase the number by an order of magnitude, the chance probability would be $\sim 4.3 \times 10^{-12}$, which is still negligibly small. Furthermore, we did not consider other constraints in the above calculation that could substantially reduce the chance probability. For example, because GN-z11 is at Dec. ~ 62º, an object with a distance ≳ 1 AU (astronomical units) to the Earth at the moment of GN-z11-flash would have a high inclination angle, which would produce an extended spectrum. It is exactly the reason that we previously ruled out a man-made satellite with height ≳4000 km. This puts a very strong constraint on the distance of a moving object and would largely reduce the above chance probability, because the vast majority of the known asteroids are at ≳ 2 AU with respect to the Sun.

Sixth, we ruled out the possibility of a foreground brown dwarf flare, or a foreground brown dwarf flare aligned with a high-redshift galaxy. GN-z11 is an extended object in HST images, so it is not a Galactic star. Deep HST and Spitzer IRAC photometry indicates that GN-z11 is a galaxy at $z > 10$, and deep HST grism spectra suggest that GN-z11 is at $z = 11.1$

± 0.1. The HST and Spitzer photometry has ruled out the probability that GN-z11 is an extreme line emitter or a red and dusty galaxy at low redshift ($z \sim 2$ - $2.5$). GN-z11's optical and near-IR colours are not consistent with the existence of a brown dwarf. Therefore, the probability that GN-z11-flash was caused by a foreground brown dwarf is zero.

Finally, our team members have a large amount of multi-slit spectroscopic data collected in the past ~20 years from Keck MOSFIRE, Keck DEIMOS, LBT LUCI, Magellan IMACS, etc. We have never seen this kind of transient event in these data. Therefore, this transient must be an extremely low-probability event.

We did not detect significant absorption lines at $z = 10.957$ in the $K$-band spectrum. This is expected due to two reasons. One reason is that our $K$-band spectrum covers a very short rest-frame wavelength range from ~1700 to ~1950 Å (excluding the range severely affected by telluric absorption features at < 2.025 μm; see Fig. 2) if the flash is at $z = 10.957$. There are no strong absorption lines in this wavelength range. The other reason is the low signal-to-noise of the spectrum, as we can see in Fig. 2.

The line width of GN-z11-flash in the spatial direction ($y$-axis) in the 2D image (e.g., Fig. 1a or Fig. 2a, where the $x$-axis indicates the wavelength and the $y$-axis represents the spatial position) appears to be slightly broader than that of the PSF, as demonstrated in Extended Data Fig. 1. In Extended Data Fig. 1a, the blue profiles represent GN-z11-flash at 10 different positions in the wavelength direction, and the black profiles represent a bright reference star (i.e., PSF) on a neighboring slit at the same 10 positions (same $x$-axis positions). Each of the profiles is the combination of 30 individual profiles to increase the signal-to-noise ratio (S/N). The profiles have also been shifted to a common center at $x = 10$ (the central positions of GN-z11-flash have a scatter of up to 1 pixel due to low S/N). We have removed the profiles that were likely affected by OH sky lines. Compared to the reference star, the GN-z11-flash line width is about 20%−25% larger. The reason for the broader line width is that GN-z11-flash was likely off the slit center during our observations, as demonstrated in Extended Data Fig. 1b. This can easily happen due to three possible reasons or the combination of some of them. The first reason is that the target positioning was done via blind acquisition using alignment stars. It is normal that some targets could be off the slit centers by 1-2 pixels in multi-object spectroscopic observations. The second reason is that GN-z11 is extended over 0.6", and the exact position of GN-z11-flash is unknown. The third reason is the possible existence of a small positional drift across the slit direction during the observations. Note that the drift along the slit direction has been clearly detected in our observations[17] and in the literature[31]. We use the nearest alignment star to show how the slit offset affects the line width for GN-z11-flash. In Extended Data Fig. 1b, the black profile presents a slit centered on the brightest pixel. Compared to the black profile, the orange and blue profiles are wider by 5% and 32%, and their flux is reduced by a factor of 1.7 and 3.4, respectively. Based on the above estimate that the GN-z11-flash line is about 20%−25% wider than the PSF, the flux loss is between 0.7 and 2.4 times the observed value. Therefore, the flux density in the $K$ band, after corrected for the slit loss, is roughly 0.1−0.2 mJy.

**The γ-ray flux upper limit.** We search for a possible GRB associated with GN-z11-flash. The transient was observable by several Na I detectors of the Gamma-ray Burst Monitor (GBM) on board the NASA Fermi mission. We search the archival continuous Time-Tagged Event data for possible γ-ray emission within a duration of about 2000 s around $T_0 =$ 2017-04-07 08:07:19.86. No significant emission above the background level was found. We also check the INTEGRAL archival data, and a null result is returned around $T_0$. There is no useful Swift BAT and HXMT data publicly available.

We extract the spectra of the GBM detectors n6, n7, and b0 between $T_0 − 50$ s and $T_0 + 50$ s and compare them with the background spectra. We employ a previously developed method[32] to calculate the source count upper limit at 90% confidence level ($S_{LL,i}$, $S_{UL,i}$) in each energy channel $i$ based on the corresponding observed counts and background counts. We then calculate the flux upper limit by fitting the power-law model to a simulated spectrum realized based on ($S_{LL,i}$, $S_{UL,i}$). We find that a power-law model is adequate to fit such simulated spectra. The $1−10^4$ keV flux upper limit (90% confidence level) is found to be $F_γ = 5.9 × 10^{-7}$ erg cm$^{-2}$ s$^{-1}$.

**Physical models.** We consider two possible physical models that may allow a high-redshift GRB producing a rest-frame UV flash in the synchrotron self-absorbed regime. The first model interprets the flash as originating from the external reverse shock (RS). For standard parameters and for all the previously known RS UV/optical flashes[5,8], the UV/optical bands are well above the cooling frequency $ν_c$, so that the observed $β_ν ≈ 1.2$ cannot be interpreted. We then explore more extreme parameters to investigate whether it is possible to allow a synchrotron self-absorbed UV/optical flash (i.e., $ν < ν_a$, where $ν_a$ is the SSA frequency). Using the formalism summarized in reference[33], we go over all the possible dynamical models (an interstellar medium or a wind medium in both the thin-shell and thick-shell regimes) in various spectral regimes [$ν_a < ν_m < ν_c$, $ν_a < ν_c < ν_m$, $ν_m < ν_a < ν_c$, $ν_c < ν_a < ν_m$, and max($ν_m$, $ν_c$) < $ν_a$] to investigate whether any self-consistent solution can be reached with $ν_a$ higher than the $K$-band frequency $ν_K$, with $ν_m$ the characteristic synchrotron frequency of the electrons for the minimum Lorentz factor. No self-consistent solution can satisfy the observed conditions $ν_a > ν_K$ and $f_ν ≈ 0.1−0.2$ mJy. All the spectral regimes under consideration either have $ν_a$ below $ν_K$ or have a flux density much brighter than the observed value. We conclude that the rest-frame UV flash GN-z11-flash did not originate from the RS region.

The second model interprets GN-z11-flash as the low-energy counterpart of γ-ray emission. For the synchrotron origin of GRB prompt emission, the SSA frequency $ν_a$ is generally written as Equation 6 of reference[14]. For a typical low-energy photon index $\hat{\alpha} = -1$ (ref. 34), which is achievable within the synchrotron model[18] with an emission radius $R = 10^{14}$ $R_{14}$ cm, one has $\beta_1 = 0$, $C_1 = 1.2$, and $C_2 \approx 1.2$. In the regime $\nu_m < \nu_a < \nu_p$ regime, where $\nu_p$ is the peak frequency, the SSA frequency can be simplified to

$$\nu_a = (2.05 \times 10^{15} \text{ Hz}) \left(\frac{f_{\nu_p}}{\text{mJy}}\right)^{2/5} \nu_{p,19}^0 \left(\frac{D_{L,28}}{1+z}\right)^{4/5} \left(\frac{\Gamma_{300}}{1+z}\right)^{2/5} B_5^{1/5} R_{14}^{-4/5}, \qquad (1)$$

where $f_{\nu_p}$ is the flux at the peak frequency, $D_{L,28}$ is the luminosity distance in units of $10^{28}$ cm, and $B$ is the co-moving magnetic field strength in units of $10^5$ G. Since no $\gamma$-rays were detected to be associated with GN-z11-flash, we can only adopt a set of typical parameters to describe the GRB. We consider an energetic long GRB with the following parameters: isotropic peak luminosity $L_{\gamma,p,iso} = 10^{53}$ erg s$^{-1}$ (which satisfies the upper limit from the non-detection), observed peak energy $E_p = 100$ keV (the rest-frame peak energy $E_{p,z} \simeq 1.2$ MeV), Lorentz factor $\Gamma = 300$, and the strength of the co-moving magnetic field strength $B = 10^5$ G at $R = 5 \times 10^{14}$ cm. The radius $R$ was estimated to be $\geq$ a few times $10^{14}$ cm (ref. 14) for $\Gamma = 300$, and we adopt $5 \times 10^{14}$ cm. For these parameters, the flux at the peak energy is $f_{\nu_p} \simeq 0.25$ mJy, and the SSA frequency is $\nu_a \simeq 1.8 \times 10^{14}$ Hz. This is slightly above the $K$-band frequency $\nu_K \simeq 1.4 \times 10^{14}$ Hz. The estimated $K$-band flux density is $f_{\nu_p}(\nu_K/\nu_a)^{\beta_\nu} \approx 0.18$ mJy, consistent with the observed value ~0.1−0.2 mJy. So, our trial parameters reach a self-consistent result.

We estimate the probability that GN-z11-flash showed up in only one 3-min exposure during our observations, if it was the prompt emission of a GRB from GN-z11 at $z = 10.957$. We assume that the durations of GRB prompt emission[5-9,36] are comparable to the durations of $\gamma$-rays T90 (the time interval in which the integrated photon counts increase from 5% to 95% of the total counts). We first build the rest-frame T90 distribution from the observed T90 distribution and the redshift distribution of known GRBs. The observed T90 distribution is taken from a recent GRB catalogue[37], and has a log-normal shape ($\mu$, $\sigma$) with a mean value of $\mu = 1.476$ (29.9 s). The redshift distribution of GRBs is from a list of long GRBs obtained via a GRB archive (https://swift.gsfc.nasa.gov/archive/grb_table.html/). Our sample consists of 332 GRBs with reliable redshift measurements, and spans a wide redshift range with a median $z \sim 1.85$. We assume that the rest-frame T90 distribution also has a log-normal shape ($\mu_0$, $\sigma_0$) (we use subscript 0 to denote rest frame), and we try a range of ($\mu_0$, $\sigma_0$) pairs with steps $\Delta\mu_0 = 0.02$ and $\Delta\sigma_0 = 0.01$. For each of the 332 GRBs, we simulate 1000 observed T90 durations for a given ($\mu_0$, $\sigma_0$) pair. We then estimate a simulated ($\mu$, $\sigma$) for this ($\mu_0$, $\sigma_0$) pair based on the 332,000 durations. We repeat it for the other ($\mu_0$, $\sigma_0$) pairs and obtain the best ($\mu_0$, $\sigma_0$) by comparing the simulated ($\mu$, $\sigma$) values with the observed ($\mu$, $\sigma$) mentioned above. We find that the best $\mu_0$ is 1.04 (~11 s), consistent with previous studies. We then convert the rest-frame T90 distribution to the observed T90 distribution at $z = 10.957$. Finally, we estimate the probability that a prompt flash only showed up in one image during our observations. Our observations had a cadence of 179 s exposure, 33 s gap (read out and setup), 179 s exposure, 33 s gap, and so on. At each second of a 212 s duration (179 s + 33 s), we simulate 1000 flashes that follow the derived T90 distribution at $z = 10.957$, and obtain 212,000 simulated flashes in total. For each flash event, we calculate flux densities in the first exposure and the following two exposures. We check whether the highest flux density is at least 10 times the second-highest flux density. The final probability is the fraction of the flashes that satisfy this criterion. We find that the final probability is ~51%, meaning that half of GRB prompt emission events would show up in only one image during our observations.

If the GN-z11-flash emission is from the same internal emission region that gives rise to the $\gamma$-rays, one would expect an afterglow following the prompt emission. Since the RS emission typically predicts a bright UV/optical flash, the non-detection during the later observations suggests that the RS emission is suppressed. This is consistent with the model if the outflow is Poynting flux-dominated[21]. The optical emission at later epochs should be dominated by the forward shock emission. The peak flux density may be estimated as[35]

$$F_{\nu,\mathrm{max}} = (1.5\ \mu\mathrm{Jy})\ \left(\frac{1+z}{11.957}\right)\ \left(\frac{E_{52}}{\mathrm{n}}\right)\ \epsilon_{B,-4}^{1/2}\ \left(\frac{D_L}{3.63\times10^{29}\ \mathrm{cm}}\right)^2,\qquad(2)$$

where $E = 10^{52}\ E_{52}$ erg is the isotropic energy of the blast wave, $n$ is the medium density in units of number per cm$^3$, $\epsilon_B$ is the fraction of shock energy that is given to magnetic fields, and $D_L$ is the luminosity distance that is normalized to the value at $z = 10.957$. One can see that this flux density is below the sensitivity limit of our Keck observations, which is consistent with the non-detection of emission after the detection of GN-z11-flash.

GN-z11 is a luminous star-forming galaxy with a UV star formation rate (SFR) of $26 \pm 3$ $M_\odot$ yr$^{-1}$ ($M_\odot$ is the solar mass). We estimate the chance probability of detecting one GRB during our observations. The local event rate densities of high-luminosity and low-luminosity long GRBs are roughly 0.8 and 160 Gpc$^{-3}$ yr$^{-1}$, respectively[38]. The local SFR density is roughly 0.02 $M_\odot$ yr$^{-1}$ Mpc$^{-3}$ (ref. 39). Applying this local relation between the GRB rate and SFR to GN-z11, we obtain a GRB rate of (~1-200) Myr$^{-1}$ per galaxy. This gives a chance probability of $(0.5 \sim 100) \times 10^{-10}$ of detecting one GRB during our observations of GN-z11. We reduce this probability to $(0.3 \sim 60) \times 10^{-10}$ by considering that ~60% of the known GRBs are as bright as GN-z11-flash in the UV/optical. In case that a significant fraction of GRB progenitors can produce luminous bursts at high redshift, the chance probability is greater than $1.0 \times 10^{-9}$. Note that our current knowledge about the GRB rate at high redshift is very limited. In the above calculation, we did not consider any possible redshift evolution of the relation between the GRB rate and galaxy properties (such as stellar mass, SFR, metallicity, dust, and so on), despite the fact that galaxy properties evolve significantly toward higher redshift.

**Data availability** The Keck MOSFIRE data of this work are publicly available from the Keck Observatory Archive (https://www2.keck.hawaii.edu/koa/public/koa.php). The long GRBs used for a simulation are publicly available from a GRB archive website (https://swift.gsfc.nasa.gov/archive/grb_table.html/). The source data for the figures within this paper are provided as Source Data files. Other data of this study are available from the corresponding authors upon reasonable request.

**Code availability** The Keck MOSFIRE data were reduced using a publicly available data reduction pipeline (https://github.com/Keck-DataReductionPipelines/MosfireDRP). The GN-z11-flash spectra were extracted and calibrated using standard IRAF routines (https://iraf-community.github.io/).

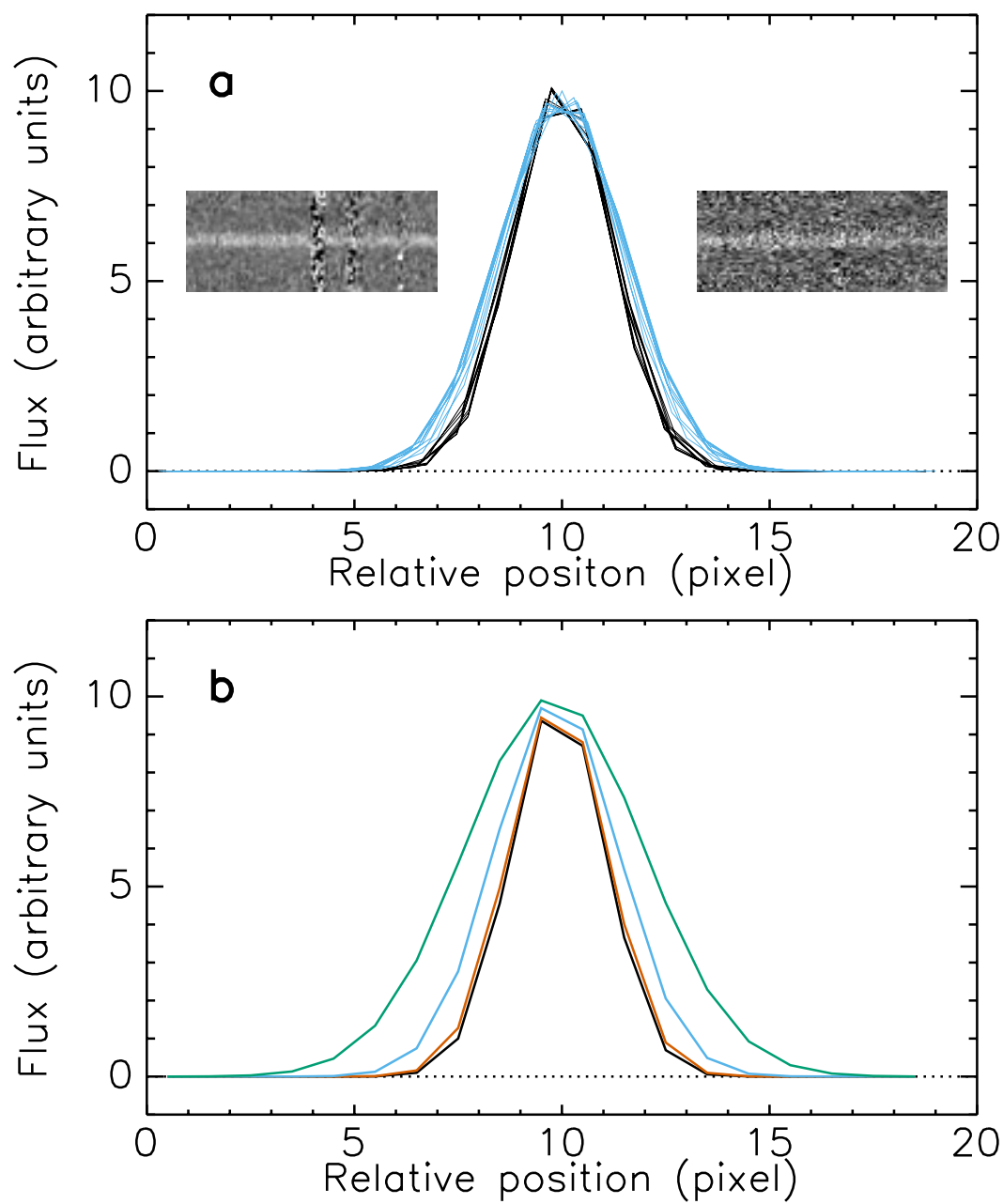

**Extended Data Fig. 1 Line width of the GN-z11-flash 2D spectrum in the spatial direction.** (a) Comparison with a reference star. Each of the 10 blue profiles (centered at $x = 10$) is the best Gaussian fit to an observed profile at one wavelength position. The observed profile is the stack of 30 individual profiles along the wavelength direction. The 10 black profiles (also centered at $x = 10$) are the best-fitted profiles at the same positions for a bright reference star on a neighboring slit. The profiles are normalized so that the peak values are about 10. The insets show parts of the 2D spectrum centered at 2.15 and 2.25 μm, respectively. (b) Line profiles of an alignment star. The black profile presents a slit centered on the brightest pixel. The orange, blue, and green profiles present three slits that are off the center by +2, +3, and +4 pixels, respectively. The profiles are normalized so that the peak values are about 10.